\documentclass[showpacs,preprintnumbers,amsmath,amssymb,prl,twocolumn,floatfix,amssymb]{revtex4}
\usepackage[dvips]{graphicx,color}
\usepackage{dcolumn}
\usepackage{amsmath,amssymb,bbold,bm}

\begin{document}
\title{Kondo Temperature in Multilevel Quantum Dots}
\author{Ion Garate$^{1,2}$ and Ian Affleck$^{1,2}$}
\affiliation{$^1$Department of Physics and Astronomy, The University of British Columbia, Vancouver, BC V6T 1Z1, Canada}
\affiliation{$^2$Canadian Institute for Advanced Research, Toronto, ON M5G 1Z8, Canada.}
\date{\today}
\begin{abstract}
We develop a general method to evaluate the Kondo temperature in a multilevel quantum dot that is weakly coupled to conducting leads.
Our theory reveals that the Kondo temperature is strongly enhanced when the intradot energy-level spacing is comparable or smaller than the charging energy.
We propose an experiment to test our result, which consists of measuring the size-dependence of the Kondo temperature.
\end{abstract}
\maketitle

{\em Introduction.---}
The Kondo effect, a many-body phenomenon that emerges from the interaction between localized and itinerant fermionic degrees of freedom, is characterized by a low-temperature infrared (IR) divergence in perturbative calculations of physical observables such as resistivity and magnetic susceptibility\cite{coleman}.
This IR divergence is controlled by an ultraviolet (UV) cutoff $\Lambda$ that appears in the expression for the Kondo temperature via 
$T_K\simeq \Lambda (J\nu)^{1/2}\exp(-1/J\nu)$,
where $J$ is the Kondo coupling and $\nu$ is the Fermi level density of states per spin for itinerant carriers.
Such expression for $T_K$ is generally valid for $\nu J<<1$ and can be derived perturbatively starting from the venerable Kondo Hamiltonian \cite{coleman}, ${\cal H}_K=\sum \xi_k c^\dagger_k c_k +\sum_{k,k'} J c^\dagger_{k} ({\boldsymbol \sigma}/2) c_{k'} \cdot {\bf S}$.
An accurate microscopic theory of $J$ and $\Lambda$ provides crucial guidance for experimental explorations of strongly correlated electron systems.
 
A precise way to quantify $T_K$ is to work with a ``first-principles'' microscopic model that reduces to ${\cal H}_K$ at energy scales below $\Lambda$.
Quite generally this first-principles Hamiltonian can be written as ${\cal H}={\cal H}_0+{\cal H}_T$, where ${\cal H}_T$ captures the hybridization between the localized and itinerant degrees of freedom.
A perturbation theory calculation of physical observables in ${\cal H}_T$ then yields hallmark Kondo-like divergences, with $\Lambda$ and $J$ unequivocally determined in terms of the microscopic parameters of ${\cal H}$. 

Perhaps the first author to successfully implement the aforementioned scheme was Haldane\cite{haldane1978}, who evaluated the magnetic susceptibility for the single-level Anderson Hamiltonian to fourth order in the hybridization amplitude $t$. 
In the local-moment regime and for an infinite bandwidth in the continuum he obtained  $J\sim 4 t^2/E_c\equiv J_0$ and $\Lambda\simeq E_c$, where $E_c$ is the Coulomb charging energy.
Over time, Haldane's formula $T_{K,{\rm single}}\simeq E_c (\nu J_0)^{1/2}\exp(-1/\nu J_0)$ has remained as the norm for the interpretation of experimental studies of the Kondo effect in quantum dots\cite{kondo exp}, even though its applicability in these devices is {\em a priori} unclear.
A primary concern regarding Haldane's formula is that it makes no reference to the multiple energy levels present in real dots.
This concern was first addressed by Inoshita {\rm et al.} \cite{inoshita1993}, who suggested that the dense energy spectrum of quantum dots should enhance $T_K$ by several orders of magnitude.
Nevertheless, no such giant enhancement has been observed\cite{kondo exp 2}.
More recently, Aleiner {\em et al.} \cite{aleiner2002} argued that, in real quantum dots with an average single-particle spacing $\delta$, the main modification from Haldane's formula should consist of replacing $\Lambda\sim E_c$ by $\Lambda\sim{\rm min}\{E_c,\delta\}$.
The conclusions of Refs.~[\onlinecite{inoshita1993},\onlinecite{aleiner2002}] rely on effective Kondo Hamiltonians, and are thus less rigorous than the ``first-principles'' approach described above.

In this paper we follow the spirit of Ref.~[\onlinecite{haldane1978}] and construct a precise theory for $T_K$ in real quantum dots that are weakly coupled to conducting leads.
We adopt the {\em Universal Hamiltonian} \cite{aleiner2002} as an appropriate ``first-principles'' model for real quantum dots, and reach results that differ qualitatively from those of Refs.~[\onlinecite{haldane1978},\onlinecite{inoshita1993},\onlinecite{aleiner2002}]. 
In the infinite bandwidth limit we conclude that $T_{K}\simeq T_{K,{\rm single}}\exp(f)$ for $f\nu J_0<<1$, where $f$ is a function of $E_c/\delta$ (Fig. 1).
This result predicts an unconventional dependence of $T_K$ on the size of the quantum dot.
\begin{figure}[h]
\begin{center}
\includegraphics[scale=0.3,angle=270]{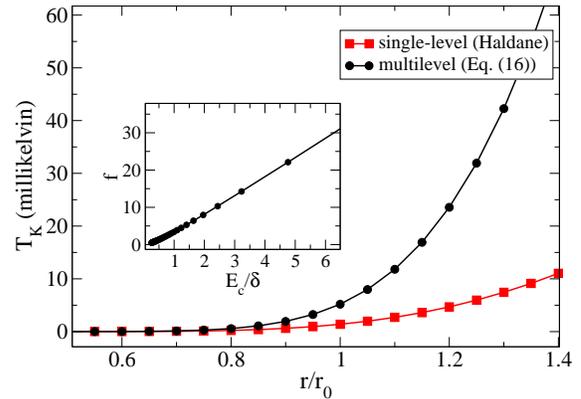}
\caption{Single-level {\em vs.} multilevel Kondo temperature for dots of variable size and fixed dot-lead tunneling rate, at the particle-hole symmetric point. $r_0$ is a lengthscale defined in the text. {\em Inset}: the function $f$ of Eq.~(\ref{eq:L_dot}) for a quantum dot with infinite equally-spaced energy levels.}
\label{fig:2}
\end{center}
\end{figure}

{\em Method.---} 
Our calculation centers on spin-flip matrix elements of an effective Hamiltonian, 
\begin{equation}
\label{eq:A}
A_{i\to f}=\langle f_0|{\cal H}_{\rm eff}|i_0\rangle,
\end{equation}
where $|i_0\rangle$ and $|f_0\rangle$ are degenerate eigenstates of ${\cal H}_0$, and ${\cal H}_{\rm eff}$ is the effective Hamiltonian derived from degenerate perturbation theory in ${\cal H}_T$. 
$|i_0\rangle$ and $|f_0\rangle$ are tensor products of a target (i.e. the localized degrees of freedom) and a projectile (i.e. an itinerant particle that scatters off the target).
Both the spin of the projectile and the spin of the target are flipped in the course of spin-flip processes.
The calculation of Eq.~(\ref{eq:A}) is considerably simpler than that of the magnetic susceptibility in Ref.~[\onlinecite{haldane1978}] because it requires neither partition functions nor external magnetic fields.
In spite of its relative simplicity, Eq.~(\ref{eq:A}) is closely connected to the scattering T-matrix and thus to a physical observable, namely the scattering rate. 

In view of the above connection, our approach exploits the long-known fact\cite{suhl1965} that in the Kondo model the spin-flip matrix elements of the T-matrix produce the ``running'' Kondo coupling
\begin{equation}
\label{eq:kondo}
J(T)=J+J^2 \nu\ln(\Lambda/T)+...,
\end{equation}
where $T$ is the temperature.
The computation of Eq.~(\ref{eq:A}) from a ``first-principles'' model and its subsequent identification with Eq.~(\ref{eq:kondo}) produces the desired explicit expression for $J$ and $\Lambda$ in terms of microscopic parameters.

{\em Effective Hamiltonian.---}
Given a Hamiltonian ${\cal H}={\cal H}_0+{\cal H}_T$, where ${\cal H}_0$ has a degenerate energy spectrum, there exists a perturbative Green's function technique\cite{messiah} to construct its exact eigen energies.
According to this approach, the key eigenvalue equation to be solved is
\begin{equation}
\label{eq:eigen}
(P_0 {\cal H}\,\, {\cal U}-E_\alpha) |E_0; \alpha\rangle=0,
\end{equation}
where $P_0$ is the projection operator onto the degenerate subspace spanned by the eigenvectors of the unperturbed energy $E_0$, $E_\alpha$ are the eigenvalues of ${\cal H}$ and $|E_0;\alpha\rangle$ are the projections of the corresponding eigenvectors onto the projected subspace. 
Hence $P_0 {\cal H}\,{\cal U}$, which is hermitian on the projected subspace, may be identified with an effective Hamiltonian. 
Also, ${\cal U}=\sum_{n=0}^\infty U^{(n)}$ and
\begin{equation}
\label{eq:Un}
U^{(n)}=\sum_{(n)}' S^{k_1} {\cal H}_T S^{k_2} {\cal H}_T...{\cal H}_T S^{k_n} {\cal H}_T P_0,
\end{equation}
for non-negative integers $k_1,...k_n$.
In Eq.~(\ref{eq:Un}), $S^k=-P_0$ if $k=0$ and $S^k=Q_0 (E_0-{\cal H}_0)^{-k} Q_0$ if $k>0$.
In addition, $Q_0={\bf 1}-P_0$.
$\sum_{(n)}'$ is extended over all sets of non-negative integers $k_1,k_2,...,k_n$ satisfying the conditions
$ k_1+k_2+...+k_p\geq p$  $(p=1,2,...,n-1)$ and $k_1+k_2+...+k_n = n$.
From Eq.~(\ref{eq:eigen}) it follows\cite{messiah} that
\begin{equation}
\label{eq:messiah}
{\cal H}_{\rm eff} \equiv P_0 {\cal H} {\cal U}=E_0 P_0 +P_0 {\cal H}_T {\cal U}.
\end{equation}

{\em Single-Level Anderson Model.---}
In order to verify that Eq.~(\ref{eq:A}) produces the correct $\Lambda$ and $J$, we employ the simplest first-principles model for which rigorous results have been long established\cite{haldane1978}.
Using the standard notation, the Anderson Hamiltonian is ${\cal H}={\cal H}_0+{\cal H}_T$, where
\begin{eqnarray}
\label{eq:anderson}
{\cal H}_0 &=& \sum_{k\sigma}\xi_k c^\dagger_{k\sigma}c_{k\sigma}+\sum_\sigma \epsilon_d d^\dagger_\sigma d_\sigma + E_c n_{d\uparrow} n_{d\downarrow}\nonumber\\
{\cal H}_T &=& t\sum_{k\sigma} (c^\dagger_{k\sigma}d_\sigma+{\rm h.c.}).
\end{eqnarray}
We evaluate $A_{i\to f}$ to fourth order in the hybridization amplitude $t$.
We choose 
$|i_0\rangle = c^\dagger_{k\uparrow} |\o\rangle d^\dagger_\downarrow|0\rangle \mbox{   and   } |f_0\rangle = c^\dagger_{k'\downarrow}|\o\rangle d^\dagger_\uparrow|0\rangle$
as the initial and final scattering states. 
$|\o\rangle$ is the Fermi sea in the continuum and $|0\rangle$ denotes the empty state of the localized level.
The momentum of the projectile is assumed to be close to the Fermi surface (i.e. $\xi_k=\xi_{k'}\simeq 0$).
$|i_0\rangle$ and $|f_0\rangle$ can be connected only via spin-flip processes; this choice is convenient in that it filters out spin-independent scattering.
 
The effective Hamiltonian may be evaluated using Eq.~(\ref{eq:messiah}) and noting that $P_0$ projects onto a two-dimensional subspace spanned by $|i_0\rangle$ and $|f_0\rangle$; the outcome reads ${\cal H}_{\rm eff}\simeq {\cal H}_0 +{\cal H}_{\rm eff}^{(2)}+{\cal H}_{\rm eff}^{(4)}$ with ${\cal H}_{\rm eff}^{(2)} = P_0 {\cal H}_T (Q_0/a) {\cal H}_T P_0$ and
\begin{eqnarray}
\label{eq:h_eff}
{\cal H}_{\rm eff}^{(4)} &=& P_0 {\cal H}_T (Q_0/a) {\cal H}_T (Q_0/a) {\cal H}_T (Q_0/a) {\cal H}_T P_0\nonumber\\
 &-& P_0 {\cal H}_T (Q_0/a^2) {\cal H}_T P_0 P_0 {\cal H}_T (Q_0/a) {\cal H}_T P_0,\\\nonumber
\end{eqnarray}
where we have exploited $\langle i_0|({\cal H}_T)^{n={\rm odd}}|f_0\rangle=0$ and defined $Q_0/a^k\equiv Q_0 (E_0-{\cal H}_0)^{-k} Q_0$.
Our ${\cal H}_{\rm eff}$ connects states with equal energy and thus contains less information than the effective Hamiltonian derived from a fourth-order Schrieffer-Wolff\cite{coleman} transformation, with which it agrees when $\xi_k=\xi_{k'}$.
At any rate, this limitation has no practical consequences because all observable properties are determined by $k$ and $k'$ located at the Fermi surface.

From Eq.~(\ref{eq:h_eff}) the lowest order contribution to $A_{i\to f}$ reads
\begin{equation}
\label{eq:t2_0}
A^{(2)}_{i\to f}=\sum_{n}\langle f_0|{\cal H}_T|n\rangle\langle n|{\cal H}_T|i_0\rangle/(E_i-E_n),
\end{equation}
where $|n\rangle$ denotes virtual intermediate states that satisfy ${\cal H}_0|n\rangle=E_{n}|n\rangle$. Also, ${\cal H}_0|i_0\rangle=E_i |i_0\rangle$ and ${\cal H}_0|f_0\rangle=E_f |f_0\rangle$ ($E_i=E_f$ as we focus on elastic scattering).
Each time ${\cal H}_T$ acts on a state it changes the number of particles by one both in the continuum and in the localized level, yet it conserves the total number of particles and the total spin.
Accordingly $|n\rangle \in\{ |\o\rangle |2\rangle,  c^\dagger_{k\uparrow}c^\dagger_{k'\downarrow}|\o\rangle |0\rangle\}$ and $A^{(2)}_{i\to f}=t^2/(\epsilon_d+E_c)-t^2/\epsilon_d$.

Next, we compute the 4th order contribution to the scattering amplitude using ${\cal H}_{\rm eff}^{(4)}$ in Eq.~(\ref{eq:h_eff}):
\begin{widetext}
\begin{equation}
\label{eq:t4}
A^{(4)}_{i\to f}=\sum_{n_1,n_2,n_3}\frac{\langle f_0|{\cal H}_T|n_1\rangle\langle n_1|{\cal H}_T|n_2\rangle\langle n_2|{\cal H}_T|n_3\rangle\langle n_3|{\cal H}_T|i_0\rangle}{(E_i-E_{n_1})(E_i-E_{n_2})(E_i-E_{n_3})}-\epsilon_2\sum_n\frac{\langle f_0|{\cal H}_T|n\rangle\langle n|{\cal H}_T|i_0\rangle}{(E_i-E_n)^2}-A^{(2)}_{i\to f} \sum_n\frac{|\langle f_0|{\cal H}_T|n\rangle|^2}{(E_i-E_n)^2} ,
\end{equation}
\end{widetext}
where $\epsilon_2=\sum_n |\langle i_0|{\cal H}_T|n\rangle|^2/(E_i-E_n)$.
The second and third terms in Eq.~(\ref{eq:t4}) were derived by inserting $|i_0\rangle\langle i_0|+|f_0\rangle\langle f_0|={\bf 1}$ between two subsequent $P_0$ operators in Eq.~(\ref{eq:h_eff}).
 In particular, the second term in Eq.~(\ref{eq:t4}) is UV divergent and plays a crucial role in ensuring that $A_{i\to f}$ remains UV finite even when the bandwidth of the continuum states is taken to infinity.
\{$n_1$,$n_2$,$n_3$\} label intermediate states, which are collected in Table I of the Supplementary Material.
Summing over all contributions and assuming an infinite bandwidth in the continuum we obtain
\begin{equation}
\label{eq:t4_and}
A^{(4)}_{i\to f}= 2 \nu t^4\left(\frac{1}{\epsilon_d}-\frac{1}{\epsilon_d+E_c}\right)^2 \ln\frac{\sqrt{-\epsilon_d(\epsilon_d+E_c)/e}}{\omega},
\end{equation} 
where $\nu$ is the Fermi surface density of states in the continuum and $\omega$ is the infrared energy cutoff. 
For the present zero-temperature calculation $\omega\simeq\xi_k$.
$A^{(2)}_{i\to f}+A^{(4)}_{i\to f}$ can be identified (modulo a factor $1/2$) with Eq.~(\ref{eq:kondo}), which yields $\Lambda=\sqrt{|\epsilon_d|(\epsilon_d+E_c)/e}$ 
and $J=2 t^2 (1/(\epsilon_d+E_c)-1/\epsilon_d)$.
These expressions agree with those of Ref.~[\onlinecite{haldane1978}].

{\em Connection with Scattering Theory.---}
Here we show that Eq.~(\ref{eq:A}) is closely linked to a physical observable.
According to standard scattering theory\cite{suhl1965,langreth1966}, the spin-dependent scattering amplitude in the Anderson model is given by
\begin{equation}
\label{eq:t_lan}
T_{S\sigma; S'\sigma'}=\langle S'|d_{\sigma'}\frac{t^2}{E+\xi_k-{\cal H}+i 0^+} d^\dagger_\sigma|S\rangle -\left(\begin{array}{c} d_{\sigma'}\leftrightarrow d^\dagger_\sigma \\ i\leftrightarrow -i\\ \xi_k\leftrightarrow -\xi_{k'}\end{array}\right),
\end{equation}
where $\sigma$ and $\sigma'$ label the spin of the projectile, $|S\rangle$ and $|S'\rangle$ are eigenstates of the full Hamiltonian ${\cal H}$ in absence of projectiles and $E$ is the exact ground state energy, i.e. ${\cal H}|S\rangle=E|S\rangle$ and ${\cal H}|S'\rangle=E|S'\rangle$.
In the local moment regime and for a large system containing an odd number of electrons $|S\rangle$ and $|S'\rangle$ are spin 1/2 ground states.
At $t=0$ the spin resides on the localized level but for $t\neq 0$ the magnetization is spatially delocalized\cite{sorensen1996}.
Below we use $S$ and $S'$ to denote the spin direction ($\Uparrow$ or $\Downarrow$) of $|S\rangle$ and $|S'\rangle$, respectively.

We evaluate spin-flip matrix elements perturbatively for the real part of Eq.~(\ref{eq:t_lan}) with $\xi_k=\xi_{k'}\simeq 0$.
It is immediate to see that the leading order contribution agrees with $A_{i\to f}^{(2)}$.
The fourth order term involves expanding $|S\rangle$, $|S'\rangle$, ${\cal H}$ and $E$ to second order in ${\cal H}_T$; the result agrees with Eq.~(\ref{eq:t4_and}).
In particular, the $O(t^2)$ (re)normalization of $|S\rangle$ and $|S'\rangle$ \cite{sakurai} coincides with the last term in Eq.~(\ref{eq:t4}).
This easily-overlooked term is essential for the correct evaluation of the T-matrix.
In sum, $A_{i\to f}={\rm Re}(T_{\Downarrow\uparrow; \Uparrow\downarrow})$ for $\xi_k=\xi_{k'}\simeq 0$.

The SU(2) symmetry of the Anderson Hamiltonian dictates ${\rm Re}(T_{S\sigma; S'\sigma'})= A_{i\to f}{\boldsymbol\sigma}_{S S'}\cdot {\boldsymbol \sigma}_{\sigma\sigma'}$, where ${\boldsymbol\sigma}$ is a vector of Pauli matrices and we ignore spin-independent scattering.
The imaginary part of the T-matrix, which quantifies the electronic scattering rate off the localized level, can then be extracted by virtue of the optical theorem:
${\rm Im} T_\sigma \propto \sum_{S;S' \sigma'} |T_{S\sigma; S'\sigma'}|^2 \simeq \sum_{S;S'\sigma'} \left[{\rm Re}\left(T_{S\sigma; S'\sigma'}\right)\right]^2 \propto J^2+2 J^3 \log(\Lambda/\omega)+...$

{\em Multilevel Quantum Dots.---}
We are now ready to evaluate $\Lambda$ and $J$ for real quantum dots via Eq.~(\ref{eq:A}).
The Universal Hamiltonian of a quantum dot that is weakly connected to a conducting lead can be written as ${\cal H}={\cal H}_0+{\cal H}_T$, where
\begin{eqnarray}
{\cal H}_0 &=& \sum_{k\sigma}\xi_k c^\dagger_{k\sigma} c_{k\sigma}+\sum_{m\sigma}\epsilon_m d^\dagger_{m\sigma} d_{m\sigma}+E_c (\hat{N}-N_g)^2\nonumber\\
{\cal H}_T &=& \sum_{k m \sigma} t_{k m}c^\dagger_{k\sigma} d_{m\sigma}+{\rm h.c.},
\end{eqnarray}
$m$ labels the discrete single-particle energy levels in the dot, $\hat{N}$ is the number operator for dot electrons, $N_g$ is the gate charge, $E_c$ is the charging energy, and we have neglected intradot exchange interactions.
 For simplicity we take $\epsilon_m=m \delta+{\rm const}$ and $t_{k m}=t$ for $\forall (k,m)$.
These simplifications are partly justified because our theory is UV-finite (see below and the Suppl. Material).

The unperturbed initial and final scattering states are 
$|i_0\rangle =  c^\dagger_{k\uparrow}|\o\rangle d^\dagger_{0\downarrow}|2 M\rangle \mbox{   ;   } |f_0\rangle = c^\dagger_{k'\downarrow}|\o\rangle d^\dagger_{0\uparrow}|2 M\rangle$,
where $|\o\rangle$ is the Fermi sea in the lead, $c^\dagger_k$ creates a projectile in the lead just above the Fermi surface, $|2 M\rangle$ is an eigenstate of the dot containing $ 2 M$ electrons and
$d^\dagger_{0\sigma}$ creates an electron in the dot at level ``0'' located immediately above the highest ($M$-th) doubly-occupied level ($m=-M,...,M$).
The unperturbed energy is $E_i=E_f=E_{FS}+\xi_k+\epsilon_0+U_{1}$, where $E_{FS}$ is the kinetic energy of the filled Fermi seas (herein $E_{\rm FS}\equiv 0$) and $\epsilon_0$ is the kinetic energy for the singly-occupied level ``0'' (tunable by a gate voltage). $U_{n}=E_c(n-1)^2$ is the Coulomb energy cost for adding $n-1$ electrons to the dot; we have chosen $N_g=2M+1$ without loss of generality by shifting all $\epsilon_m$ by a constant.

We begin by recognizing that ${\cal H}_{\rm eff}^{(1)}={\cal H}_{\rm eff}^{(3)}=0$ and that Eq.~(\ref{eq:h_eff}) remains valid.
Therefore $A_{i\to f}^{(2)}$ and $A_{i\to f}^{(4)}$ are given by Eqs.~(\ref{eq:t2_0}) and ~(\ref{eq:t4}), respectively.
For the former we find 
\begin{equation}
\label{eq:a2d}
A^{(2)}_{i\to f}=t^2\left(\frac{1}{\epsilon_0+U_{2,1}}-\frac{1}{\epsilon_0+U_{1,0}}\right),
\end{equation}
where we used $|n\rangle\in\{|\o\rangle d^\dagger_{0\downarrow} d^\dagger_{0\uparrow}|2 M\rangle, c^\dagger_{k\uparrow} c^\dagger_{k'\downarrow}|\o\rangle |2 M\rangle\}$ and defined  $U_{i,j}\equiv U_i-U_j$.

Next, we focus on $A_{i\to f}^{(4)}$.
Its computation requires considering numerous sets of intermediate states; these are listed in Tables II, III and IV of the Supplementary Material.
For simplicity we start by separating out the contribution from the $m=0$ (singly occupied) level in the dot. 
Assuming an infinite bandwidth in the lead we arrive at
\begin{equation}
\label{eq:m0_3}
A^{(4)}_{i\to f}|_{m=0} = 2 \nu\left(\frac{t^2}{\epsilon_0+U_{2,1}}-\frac{t^2}{\epsilon_0+U_{1,0}}\right)^2\ln\frac{\Lambda_0}{\omega},
\end{equation}
where  $\Lambda_0=\sqrt{|\epsilon_0+U_{1,0}|(\epsilon_0+U_{2,1})/e}$. 
Eqs.~(\ref{eq:a2d}) and ~(\ref{eq:m0_3}) are independent of $\delta$ and essentially identical to those of the single-level Anderson model.

Finally, we sum the contributions from $m\neq 0$ levels.
These depend on $\delta$ and encode the influence of the multilevel energy spectrum in the Kondo physics. 
Tables II and III show that individual virtual processes involving $m\neq 0$ levels are plagued with IR and UV divergences.
Remarkably, different divergences end up cancelling one another, partly assisted by the last two terms in Eq.~(\ref{eq:t4}). 
On one hand, the cancellation of $m\neq 0$ infrared divergences corroborates that Kondo correlations arise only from processes involving the singly occupied level in the dot.
On the other hand, the cancellation of $m\neq 0$ ultraviolet divergences confirms that high-energy excited states in the dot and lead do not alter the physics of the Kondo effect. 
In spite of being divergence free, the influence of $m\neq 0$ levels is important and makes the Kondo coupling $\delta$-dependent.
In the infinite bandwidth limit and in proximity to the particle-hole symmetric point ($\epsilon_0\simeq 0$) we obtain
\begin{equation}
\label{eq:A4}
A^{(4)}_{i\to f}|_{m\neq 0}=\nu \frac{8 t^4}{E_c^2} \left[f +O\left(\frac{\epsilon_0^2}{E_c^2}\right)+...\right],
\end{equation}
where $f$ is a dimensionless function of $E_c/\delta$ evaluated numerically (Fig. 1 and Suppl. Material).
When $E_c<<\delta$, $f(E_c/\delta)\to 0$ and multilevel effects are negligible; in the opposite limit $f(E_c/\delta)\to 5.5 E_c/\delta>>1$ and multilevel effects are important. 

The sum of Eqs.~(\ref{eq:a2d}),~(\ref{eq:m0_3}) and (\ref{eq:A4}) can be arranged as $A_{i\to f}\simeq 1/2 (J+J^2 \nu \ln (\Lambda/\omega))$. 
For $\epsilon_0\to 0$ we obtain 
\begin{eqnarray}
\label{eq:L_dot}
\Lambda&\simeq&\delta \mbox{  ,   } J\simeq J_0\left(1+f\nu J_0+\nu J_0\ln\frac{E_c}{\delta}\right) \mbox{  ; if $E_c>\delta$}\nonumber\\
\Lambda&\simeq& E_c \mbox{  ,   } J\simeq J_0\left(1+f\nu J_0\right) \mbox{  ; if $E_c<\delta$},
\end{eqnarray}
where $J_0=(4/\pi)\Gamma/\nu E_c$ is the Kondo coupling corresponding to a single-level dot and
$\Gamma=\pi \nu t^2$ is the width of the energy levels in the dot. 
Eq.~(\ref{eq:L_dot}) is valid for $(f\nu J_0,\nu J_0)<<1$, i.e. $\Gamma<<{\rm min}(E_c,\delta)$, and constitutes the main result of this paper.
We selected $\Lambda$ on physical grounds so that it sets the energy scale below which (i) the Universal Hamiltonian maps onto the Kondo Hamiltonian, (ii) 
the renormalization group flow for $J$ is that of the simple Kondo model.

{\em Experimental Implications.---}
From Eq.~(\ref{eq:L_dot}), the Kondo temperature for a multilevel quantum dot is $T_K\simeq E_c (\nu J_0)^{1/2}\exp(-1/\nu J_0)\exp(f)$, for any $E_c/\delta$ insofar as $\Gamma<<{\rm min}(E_c,\delta)$ (this condition implies that the broadening of the many-body energy eigenvalues of the isolated dot is much smaller than the energy spacing between them). 
Fig. 1 displays $T_K$ as a function of the linear dot dimension $r$. 
Introducing a lengthscale $r_0$ such that $E_c(r_0)\equiv E_{c,0}$ and $\delta(r_0)\equiv\delta_0$, it follows that $E_c(r)\simeq E_{c0} r_0/r$ and $\delta(r)\simeq\delta_0 r_0^2/r^2$.
$\Gamma=0.1 {\rm meV} $ is kept fixed (independent of $r$) and we take $E_{c0}=1 {\rm meV}$ and $\delta_0=2 {\rm meV}$; these are reasonable extrapolations based on available experimental data.
Clearly Haldane's single-level formula is accurate for smallest dots with $E_c<<\delta$; in contrast, the multilevel enhancement of the Kondo temperature becomes important for larger dots with $E_c\gtrsim \delta$.
For $E_c/\delta>>1$, $f$ is so large that $f\nu J_0<<1$ is possible only for a very small value of $\nu J_0$, which in turn results in an unmeasurably low $T_K$.
Therefore Eq.~(\ref{eq:L_dot}) is experimentally relevant for dots with $E_c\simeq\delta$, wherein the multilevel enhancement is more modest yet still noticeable ($f\sim O(1)$) .

In conclusion, we have developed a method to evaluate the Kondo temperature of real quantum dots with unprecedented precission.
Our theory predicts an unconventional and potentially measurable size-dependence of $T_K$ in dots with $E_c\simeq \delta$.
Our formalism is valid and our results readily generalizable for models that incorporate energy-dependence in the dot-lead tunneling amplitude as well as non-uniform distribution of energy levels in the dot.

{\em Acknowledgements.--}
We are indebted to A. Andreev, O. Entin-Wohlman, J. Folk and L. Glazman for helpful conversations.
This research has been supported by NSERC and CIfAR.

\begin{widetext}

\appendix
\section{Supplementary Material}
In this supplementary section we present a list of virtual processes that contribute to the first term of Eq.(9) in the main text.
Table I corresponds to the single-level Anderson model, whereas Tables II-IV dwell on the Universal Hamiltonian.
In addition, we present the dimensionless integral that gives rise to $f$ (defined by Eq. 15 in the main text):
\begin{eqnarray}
\label{eq:f}
f &=& 2 \int_0^\infty dy \sum_{m=1}^\infty\frac{4+x_m(3+x_m)^2+9 y+x_m y (16+5 x_m)+ (6+5 x_m) y^2 +y^3}{(1+x_m)(1+y)(x_m+y)(1+x_m+y)^2(4+x_m+y)}\nonumber\\
 &=& -2 \sum_{m=1}^\infty\frac{6+9(1+x_m)\ln x_m -\frac{2(3+5 x_m+ 5 x_m^2) \ln(1+x_m)}{x_m}+\frac{x_m(4+x_m)\ln(4+x_m)}{3+x_m}}{9(1+x_m)},
\end{eqnarray}
where $x_m\equiv\epsilon_m/\delta=m\delta/E_c$ and $y\equiv\xi/E_c$.
 Eq.~(\ref{eq:f}) can be derived by adding the $80$ transition amplitudes of Tables II-IV along with the second and third term of Eq.(9) in the main text, with $m\neq 0$.
The derivation is simplified by exploiting time-reversal as well as particle-hole symmetry, although similar integrals may be derived in absence of particle-hole symmetry.
Note that the sum and integral in Eq.~(\ref{eq:f}) are UV-finite, even though numerous individual amplitudes in Tables II-IV are UV-divergent; the delicate cancellation between different UV divergences adds considerable confidence on the veracity of our results.
Moreover, we find that the main contribution to $f$ originates from states with $|\epsilon_m|\lesssim E_c$ and $|\xi|\lesssim E_c$.
These observations together justify our assumption of energy-independent tunneling amplitudes and uniform energy-level spacings.
In other words, our assumptions hold provided that the tunneling-amplitudes and the energy-level spacings vary slowly on energy scales of order $E_c$, which is typically much smaller than the Fermi energy.

\begin{table*}[h]
\caption{Virtual elastic processes to fourth order in single-particle tunneling, for the Anderson Hamiltonian. The initial and final states are in the truncated, low-energy Hilbert space whereas the intermediate states trespass into the high-energy sector. We assume particle-hole symmetry in the continuum, i.e. $\sum_q\Theta(\xi_q) F(\xi_q)=\sum_q\Theta(-\xi_q) F(-\xi_q)$ for any function $F$. We exclude intermediate states that lead to $E_i-E_{n}=0$. The $1/\xi_q$ factors contain an implicit infrared cutoff that equals the energy of the projectile ($\xi_k$). For explicit calculations we substitute $\sum_q \to \nu\int d\xi$.}
\vspace{0.1 in}
\begin{tabular}{c c c c c }
\hline\hline
Label & $|n_1\rangle$ & $|n_2\rangle$ & $|n_3\rangle$ & Contribution to $A^{(4)}_{i\to f}$ (in units of $t^4$)\\
\hline\hline
1 & $|\o\rangle |2\rangle$ & $c^\dagger_{q\downarrow}|\o\rangle|1,\uparrow\rangle$ & $|\o\rangle|2\rangle$ & $\frac{1}{(\epsilon_d+U)^2}\sum_q\frac{\Theta(\xi_q)}{\xi_q}$\\[1ex]
\hline
2 & $|\o\rangle |2\rangle$ & $c^\dagger_{q\uparrow}|\o\rangle|1,\downarrow\rangle$ & $|\o\rangle |2\rangle$ & $\frac{1}{(\epsilon_d+U)^2}\sum_q\frac{\Theta(\xi_q)}{\xi_q}$\\[1ex]
\hline
3 & $c^\dagger_{k'\downarrow}c^\dagger_{k\uparrow}|\o\rangle |0\rangle$ & $c^\dagger_{k'\downarrow}c^\dagger_{k\uparrow} c_{q\uparrow}|\o\rangle|1,\uparrow\rangle$ & $c^\dagger_{k'\downarrow}c^\dagger_{k\uparrow}|\o\rangle |0\rangle$ & $\frac{1}{\epsilon_d^2}\sum_q\frac{\Theta(-\xi_q)}{-\xi_q}$\\[1ex]
\hline
4 & $c^\dagger_{k'\downarrow}c^\dagger_{k\uparrow}|\o\rangle |0\rangle$ & $c^\dagger_{k'\downarrow}c^\dagger_{k\uparrow} c_{q\downarrow}|\o\rangle|1,\downarrow\rangle$ & $c^\dagger_{k'\downarrow}c^\dagger_{k\uparrow}|\o\rangle |0\rangle$ & $\frac{1}{\epsilon_d^2}\sum_q\frac{\Theta(-\xi_q)}{-\xi_q}$\\[1ex]
\hline
5 & $|\o\rangle |2\rangle$ & $c^\dagger_{q\downarrow}|\o\rangle|1,\uparrow\rangle$ & $c^\dagger_{q\downarrow} c^\dagger_{k\uparrow}|\o\rangle|0\rangle$ & $\frac{1}{\epsilon_d+U}\sum_q\frac{\Theta(\xi_q)}{\xi_q}\frac{1}{-\epsilon_d+\xi_q}$\\[1ex]
\hline
6 & $c^\dagger_{q\uparrow} c^\dagger_{k'\downarrow}|\o\rangle |0\rangle$ & $c^\dagger_{q\uparrow}|\o\rangle|1,\downarrow\rangle$ & $|\o\rangle |2\rangle$ & $\frac{1}{\epsilon_d+U}\sum_q\frac{\Theta(\xi_q)}{\xi_q}\frac{1}{-\epsilon_d+\xi_q}$\\[1ex]
\hline
7 & $c^\dagger_{k'\downarrow} c_{q\downarrow} |\o\rangle |2\rangle$ & $c^\dagger_{k'\downarrow} c^\dagger_{k\uparrow} c_{q\downarrow}|\o\rangle|1,\downarrow\rangle$ & $c^\dagger_{k'\downarrow} c^\dagger_{k\uparrow}  |\o\rangle |0\rangle$ & $\sum_q \frac{1}{\epsilon_d+U-\xi_q}\frac{1}{-\epsilon_d}\frac{\Theta(-\xi_q)}{-\xi_q}$\\[1ex]
\hline
8 & $c^\dagger_{k\uparrow}  c^\dagger_{k'\downarrow} |\o\rangle |0\rangle$ & $c^\dagger_{k\uparrow} c^\dagger_{k'\downarrow}  c_{q\uparrow}|\o\rangle|1,\uparrow\rangle$ & $|c^\dagger_{k\uparrow} c_{q\uparrow}  |\o\rangle |2\rangle$ & $\sum_q \frac{1}{\epsilon_d+U-\xi_q}\frac{1}{-\epsilon_d}\frac{\Theta(-\xi_q)}{-\xi_q}$\\[1ex]
\hline\hline
\end{tabular}
\end{table*}


\begin{table*}[t]
\caption{Virtual elastic processes to fourth order in single-particle tunneling, for the Universal Hamiltonian. 
We assume particle-hole symmetry in the lead and exclude intermediate states with $E_i-E_{n}=0$. Moreover we ignore the particular instances in which $q=k,k'$; these do not lead to any IR divergences and their UV divergences should cancel in the same manner as for the Anderson Hamiltonian. For explicit calculations we substitute $\sum_q \to \nu\int d\xi$.}
\vspace{0.1 in}
\begin{tabular}{ c c c c c }
\hline\hline
Label & $|n_1\rangle$ & $|n_2\rangle$ & $|n_3\rangle$ & Contribution to $A^{(4)}_{i\to f}$ (in units of $t^4$)\\
\hline\hline
1& $|\o\rangle d^\dagger_{0\uparrow} d^\dagger_{m\downarrow}|2 M\rangle$ & $c^\dagger_{q\downarrow}|\o\rangle d^\dagger_{0\uparrow}|2 M\rangle$ & $|\o\rangle d^\dagger_{0\uparrow} d^\dagger_{0\downarrow} |2 M\rangle$ &  $\frac{1}{\epsilon_0+U_{2}-U_{1}}\sum\limits_{m\geq 0;q}\frac{1}{\epsilon_m+U_{2}-U_{1}}\frac{\Theta(\xi_q)}{\xi_q}$ \\[1ex]
\hline
2& $|\o\rangle d^\dagger_{0\uparrow} d^\dagger_{0\downarrow} |2M\rangle$ & $c^\dagger_{q\uparrow}|\o\rangle d^\dagger_{0\downarrow}|2 M\rangle$ & $|\o\rangle d^\dagger_{0\downarrow} d^\dagger_{m\uparrow}|2 M\rangle$ & same as previous\\[1ex]
\hline
3& $c^\dagger_{k\uparrow} c^\dagger_{k'\downarrow}|\o\rangle|2 M\rangle$ & $c^\dagger_{k\uparrow} c^\dagger_{k'\downarrow} c_{q\uparrow}|\o\rangle d^\dagger_{m\uparrow}|2M\rangle$ & $c^\dagger_{k\uparrow} c^\dagger_{k'\downarrow}|\o\rangle|2 M\rangle$ & $\frac{1}{(\epsilon_0+U_{1}-U_{0})^2}\sum\limits_{m\geq 0 ; q}\frac{\Theta(\xi_q)}{\epsilon_m-\epsilon_0+\xi_q}$  \\[1ex]
\hline
4& $c^\dagger_{k\uparrow} c^\dagger_{k'\downarrow}|\o\rangle|2 M\rangle$ & $c^\dagger_{k\uparrow} c^\dagger_{k'\downarrow} c_{q\downarrow}|\o\rangle d^\dagger_{m\downarrow}|2M\rangle$ & $c^\dagger_{k\uparrow} c^\dagger_{k'\downarrow}|\o\rangle|2 M\rangle$ &  same as previous \\[1ex]
\hline
5& $|\o\rangle d^\dagger_{0\uparrow} d^\dagger_{m\downarrow}|2M\rangle$ & $c^\dagger_{q\downarrow}|\o\rangle d^\dagger_{0\uparrow}|2M\rangle$ & $c^\dagger_{q\downarrow}c^\dagger_{k\uparrow}|\o\rangle|2M\rangle$ & $\sum\limits_{m\geq 0 ; q}\frac{1}{\epsilon_m+U_{2}-U_{1}}\frac{\Theta(\xi_q)}{\xi_q}\frac{1}{\xi_q-\epsilon_0+U_{0}-U_{1}}$ \\[1ex]
\hline
6& $c^\dagger_{q\uparrow}c^\dagger_{k'\downarrow}|\o\rangle|2 M\rangle$ & $c^\dagger_{q\uparrow}|\o\rangle d^\dagger_{0\downarrow}|2M\rangle$ & $|\o\rangle d^\dagger_{0\downarrow}d^\dagger_{m\uparrow}|2M\rangle$ & same as previous   \\[1ex]
\hline
7& $c^\dagger_{k'\downarrow} c_{q\downarrow}|\o\rangle d^\dagger_{0\uparrow} d^\dagger_{m\downarrow}|2M\rangle$ & $c^\dagger_{k'\downarrow}c^\dagger_{k\uparrow} c_{q\downarrow}|\o\rangle d^\dagger_{m\downarrow}|2M\rangle$ & $c^\dagger_{k'\downarrow} c^\dagger_{k\uparrow}|\o\rangle|2M\rangle$ & $\frac{1}{-\epsilon_0+U_{0}-U_{1}}\sum\limits_{m\geq 0 ; q}\frac{\Theta(\xi_q)}{\xi_q+\epsilon_m-\epsilon_0}\frac{1}{\xi_q+\epsilon_m+U_{2}-U_{1}}$  \\[1ex]
\hline
8& $c^\dagger_{k\uparrow} c^\dagger_{k'\downarrow}|\o\rangle|2M\rangle$ & $c^\dagger_{k\uparrow} c^\dagger_{k'\downarrow} c_{q\uparrow}|\o\rangle d^\dagger_{m\uparrow}|2M\rangle$ & $c^\dagger_{k\uparrow}c_{q\uparrow}|\o\rangle d^\dagger_{0\downarrow} d^\dagger_{m\uparrow}|2M\rangle$ &  same as previous \\[1ex]
\hline
9& $|\o\rangle d^\dagger_{0\uparrow} d^\dagger_{m\downarrow}|2M\rangle$ & $c_{q\downarrow}|\o\rangle d^\dagger_{0\uparrow} d^\dagger_{0\downarrow} d^\dagger_{m\downarrow}|2M\rangle$ & $|\o\rangle d^\dagger_{0\downarrow} d^\dagger_{0\uparrow}|2M\rangle$ & $\frac{-1}{\epsilon_0+U_{2}-U_{1}}\sum\limits_{m>0 ; q}\frac{1}{\epsilon_m+U_{2}-U_{1}}\frac{\Theta(\xi_q)}{\xi_q+\epsilon_0+\epsilon_m+U_{3}-U_{1}}$  \\[1ex]
\hline
10& $|\o\rangle d^\dagger_{0\uparrow} d^\dagger_{0\downarrow}|2M\rangle$ & $c_{q\uparrow}|\o\rangle d^\dagger_{0\uparrow} d^\dagger_{0\downarrow} d^\dagger_{m\uparrow}|2M\rangle$ & $|\o\rangle d^\dagger_{0\downarrow} d^\dagger_{m\uparrow}|2M\rangle$ &  same as previous \\[1ex]
\hline
11& $|\o\rangle d^\dagger_{0\uparrow} d^\dagger_{0\downarrow}|2M\rangle$ & $c_{q\downarrow}|\o\rangle d^\dagger_{0\uparrow} d^\dagger_{0\downarrow} d^\dagger_{m\downarrow}|2M\rangle$ & $|\o\rangle d^\dagger_{0\downarrow} d^\dagger_{0\uparrow}|2M\rangle$ &  $\frac{1}{(\epsilon_0+U_{2}-U_{1})^2}\sum\limits_{m>0 ; q}\frac{\Theta(\xi_q)}{\xi_q+\epsilon_0+\epsilon_m+U_{3}-U_{1}}$ \\[1ex]
\hline
12& $|\o\rangle d^\dagger_{0\uparrow} d^\dagger_{0\downarrow}|2M\rangle$ & $c_{q\uparrow}|\o\rangle d^\dagger_{0\uparrow} d^\dagger_{0\downarrow} d^\dagger_{m\uparrow}|2M\rangle$ & $|\o\rangle d^\dagger_{0\downarrow} d^\dagger_{0\uparrow}|2M\rangle$ &  same as previous \\[1ex]
\hline
13& $|\o\rangle d^\dagger_{0\uparrow} d^\dagger_{0\downarrow}|2M\rangle$ & $c_{q\uparrow}|\o\rangle d^\dagger_{0\uparrow} d^\dagger_{0\downarrow} d^\dagger_{m\uparrow}|2M\rangle$ & $c_{q\uparrow}c^\dagger_{k\uparrow}|\o\rangle d^\dagger_{0\downarrow} d^\dagger_{0\uparrow}|2M\rangle$ & $-\frac{1}{\epsilon_0+U_{2}-U_{1}}\sum\limits_{m>0; q}\frac{\Theta(\xi_q)}{\xi_q+\epsilon_0+U_{2}-U_{1}}\frac{1}{\xi_q+\epsilon_0+\epsilon_m+U_{3}-U_{1}}$   \\[1ex]
\hline
14& $c_{q\downarrow} c^\dagger_{k'\downarrow}|\o\rangle d^\dagger_{0\uparrow} d^\dagger_{0\downarrow}|2M\rangle$ & $c_{q\downarrow}|\o\rangle d^\dagger_{0\uparrow} d^\dagger_{0\downarrow} d^\dagger_{m\downarrow}|2M\rangle$ & $|\o\rangle d^\dagger_{0\downarrow} d^\dagger_{0\uparrow}|2M\rangle$ &  same as previous \\[1ex]
\hline
15& $c^\dagger_{k\uparrow} c^\dagger_{k'\downarrow}|\o\rangle|2M\rangle$ & $c^\dagger_{k\uparrow} c^\dagger_{k'\downarrow} c^\dagger_{q\uparrow}|\o\rangle d_{m\uparrow}|2M\rangle$ & $c^\dagger_{k\uparrow}c^\dagger_{k'\downarrow}|\o\rangle|2M\rangle$ &  $\frac{1}{(\epsilon_0+U_{1}-U_{0})^2}\sum\limits_{m<0 ; q}\frac{\Theta(\xi_q)}{\xi_q-\epsilon_m-\epsilon_0+U_{-1}-U_{1}}$ \\[1ex]
\hline
16& $c^\dagger_{k\uparrow} c^\dagger_{k'\downarrow}|\o\rangle|2M\rangle$ & $c^\dagger_{k\uparrow} c^\dagger_{k'\downarrow} c^\dagger_{q\downarrow}|\o\rangle d_{m\downarrow}|2M\rangle$ & $c^\dagger_{k\uparrow}c^\dagger_{k'\downarrow}|\o\rangle|2M\rangle$ &  same as previous \\[1ex]
\hline
17& $c^\dagger_{q\uparrow} c^\dagger_{k'\downarrow}|\o\rangle|2M\rangle$ & $c^\dagger_{k\uparrow} c^\dagger_{k'\downarrow} c^\dagger_{q\uparrow}|\o\rangle d_{m\uparrow}|2M\rangle$ & $c^\dagger_{k\uparrow}c^\dagger_{k'\downarrow}|\o\rangle|2M\rangle$ &  $\frac{1}{\epsilon_0+U_{1}-U_{0}}\sum\limits_{m<0 ; q}\frac{\Theta(\xi_q)}{\xi_q-\epsilon_0+U_{0}-U_{1}}\frac{1}{\xi_q-\epsilon_m-\epsilon_0+U_{-1}-U_{1}}$ \\[1ex]
\hline
18& $c^\dagger_{k\uparrow} c^\dagger_{k'\downarrow}|\o\rangle|2M\rangle$ & $c^\dagger_{k\uparrow} c^\dagger_{k'\downarrow} c^\dagger_{q\downarrow}|\o\rangle d_{m\downarrow}|2M\rangle$ & $c^\dagger_{k\uparrow}c^\dagger_{q\downarrow}|\o\rangle|2M\rangle$ &  same as previous \\[1ex]
\hline
19& $|\o\rangle d^\dagger_{0\uparrow} d^\dagger_{0\downarrow}|2M\rangle$ & $c^\dagger_{q\downarrow}|\o\rangle d^\dagger_{0\uparrow} d^\dagger_{0\downarrow} d_{m\downarrow}|2M\rangle$ & $|\o\rangle d^\dagger_{0\uparrow} d^\dagger_{0\downarrow} |2M\rangle$ &  $\frac{1}{(\epsilon_0+U_{2}-U_{1})^2}\sum\limits_{m<0 ; q}\frac{\Theta(\xi_q)}{\xi_q-\epsilon_m+\epsilon_0}$  \\[1ex]
\hline
20& $|\o\rangle d^\dagger_{0\uparrow} d^\dagger_{0\downarrow}|2M\rangle$ & $c^\dagger_{q\uparrow}|\o\rangle d^\dagger_{0\uparrow} d^\dagger_{0\downarrow} d_{m\uparrow}|2M\rangle$ & $|\o\rangle d^\dagger_{0\uparrow} d^\dagger_{0\downarrow} |2M\rangle$ &  same as previous  \\[1ex]
\hline
21& $c^\dagger_{k'\downarrow} c_{q\downarrow} |\o\rangle d^\dagger_{0\uparrow} d^\dagger_{0\downarrow} |2M\rangle$ & $c^\dagger_{k'\downarrow}c^\dagger_{k\uparrow} c_{q\downarrow}|\o\rangle d^\dagger_{0\downarrow}|2M\rangle$ & $c^\dagger_{k'\downarrow} c^\dagger_{k\uparrow} |\o\rangle d^\dagger_{0\downarrow}d_{m\downarrow}|2M\rangle$  & $\sum\limits_{m<0 ; q}\frac{1}{\epsilon_m+U_{1}-U_{0}}\sum_q\frac{\Theta(\xi_q)}{\xi_q}\frac{1}{-\xi_q-\epsilon_0+U_{1}-U_{2}}$ \\[1ex]
\hline
22& $c^\dagger_{k'\downarrow} c^\dagger_{k\uparrow} |\o\rangle d^\dagger_{0\uparrow} d_{m\uparrow} |2M\rangle$ & $c^\dagger_{k'\downarrow}c^\dagger_{k\uparrow} c_{q\uparrow}|\o\rangle d^\dagger_{0\uparrow}|2M\rangle$ & $c^\dagger_{k\uparrow} c_{q\uparrow} |\o\rangle d^\dagger_{0\downarrow}d^\dagger_{0\uparrow}|2M\rangle$  &  same as previous\\[1ex]
\hline
23& $c^\dagger_{k'\downarrow} c_{q\uparrow}|\o\rangle d^\dagger_{0\uparrow} d^\dagger_{m\uparrow}|2M\rangle$ & $c^\dagger_{k'\downarrow} c^\dagger_{k\uparrow} c_{q\uparrow}|\o\rangle d^\dagger_{0\uparrow}|2M\rangle$ & $c^\dagger_{k\uparrow} c_{q\uparrow}|\o\rangle d^\dagger_{0\uparrow} d^\dagger_{0\downarrow}|2M\rangle$ & $-\sum\limits_{m>0 ; q}\frac{\Theta(\xi_q)}{\xi_q}\frac{1}{\xi_q+\epsilon_m+U_{2}-U_{1}}\frac{1}{\xi_q+\epsilon_0+U_{2}-U_{1}}$   \\[1ex]
\hline
24& $c^\dagger_{k'\downarrow} c_{q\downarrow}|\o\rangle d^\dagger_{0\uparrow} d^\dagger_{0\downarrow}|2M\rangle$ & $c^\dagger_{k'\downarrow} c^\dagger_{k\uparrow} c_{q\downarrow}|\o\rangle d^\dagger_{0\downarrow}|2M\rangle$ & $c^\dagger_{k\uparrow} c_{q\downarrow}|\o\rangle d^\dagger_{0\downarrow} d^\dagger_{m\downarrow}|2M\rangle$ &  same as previous  \\[1ex]
\hline
25& $|\o\rangle d^\dagger_{0\uparrow}d^\dagger_{0\downarrow}|2M\rangle$ & $c^\dagger_{q\uparrow}|\o\rangle d^\dagger_{0\downarrow}|2M\rangle$ & $c^\dagger_{q\uparrow} c^\dagger_{k\uparrow}|\o\rangle d^\dagger_{0\downarrow}d_{m\uparrow}|2M\rangle$ &  $-\frac{1}{-\epsilon_0+U_{1}-U_{2}}\sum\limits_{m<0 ; q}\frac{\Theta(\xi_q)}{\xi_q}\frac{1}{\epsilon_m+U_{1}-U_{0}-\xi_q}$   \\[1ex]
\hline
26& $c^\dagger_{k'\downarrow}c^\dagger_{q\downarrow}|\o\rangle d^\dagger_{0\uparrow}d_{m\downarrow}|2M\rangle$ & $c^\dagger_{q\downarrow}|\o\rangle d^\dagger_{0\uparrow}|2M\rangle$ & $|\o\rangle d^\dagger_{0\downarrow}d^\dagger_{0\uparrow}|2M\rangle$ & same as previous    \\[1ex]
\hline
27& $c^\dagger_{k'\downarrow} c_{q\uparrow}|\o\rangle d^\dagger_{0\uparrow}d^\dagger_{m\uparrow}|2M\rangle$ & $c^\dagger_{k'\downarrow} c^\dagger_{k\uparrow} c_{q\uparrow}|\o\rangle d^\dagger_{0\uparrow}|2M\rangle$  &  $c^\dagger_{k'\downarrow} c^\dagger_{k\uparrow} |\o\rangle|2M\rangle$ & $-\frac{1}{-\epsilon_0+U_{0}-U_{1}}\sum\limits_{m>0 ; q}\frac{\Theta(\xi_q)}{\xi_q}\frac{1}{\xi_q+\epsilon_m+U_{2}-U_{1}}$   \\[1ex]
\hline
28& $c^\dagger_{k'\downarrow} c^\dagger_{k\uparrow}|\o\rangle|2M\rangle$ & $c^\dagger_{k'\downarrow} c^\dagger_{k\uparrow} c_{q\downarrow}|\o\rangle d^\dagger_{0\downarrow}|2M\rangle$ & $c^\dagger_{k\uparrow} c_{q\downarrow} |\o\rangle d^\dagger_{0\downarrow} d^\dagger_{m\downarrow}|2M\rangle$  &  same as previous \\[1ex]
\hline
29 & $c^\dagger_{k'\downarrow} c^\dagger_{k\uparrow}|\o\rangle d^\dagger_{0\uparrow} d_{m\uparrow}|2 M\rangle$ & $c^\dagger_{k'\downarrow} c^\dagger_{k\uparrow} c_{q\uparrow}|\o\rangle d^\dagger_{0\uparrow}|2 M\rangle $ & $c^\dagger_{k'\downarrow} c^\dagger_{k\uparrow} |\o\rangle| 2 M\rangle$ & $\frac{1}{\epsilon_0+U_1-U_0}\sum\limits_{m<0 ; q}\frac{1}{\epsilon_m+U_1-U_0}\frac{\Theta(\xi_q)}{\xi_q}$   \\[1ex]
\hline
30 & $c^\dagger_{k'\downarrow} c^\dagger_{k\uparrow}|\o\rangle|2 M\rangle$ & $c^\dagger_{k'\downarrow} c^\dagger_{k\uparrow} c_{q\downarrow}|\o\rangle d^\dagger_{0\downarrow}|2 M\rangle $ & $c^\dagger_{k'\downarrow} c^\dagger_{k\uparrow} |\o\rangle d^\dagger_{0\downarrow} d_{m\downarrow}|2 M\rangle$ & same as previous \\[1ex]
\hline
31 & $c^\dagger_{k'\downarrow} c^\dagger_{q\uparrow}|\o\rangle|2 M\rangle$ & $c^\dagger_{q\uparrow}|\o\rangle d^\dagger_{0\downarrow}|2M\rangle$ & $c^\dagger_{q\uparrow} c^\dagger_{k\uparrow}|\o\rangle d^\dagger_{0\downarrow} d_{m\uparrow}|2 M\rangle$ & $-\sum\limits_{m<0 ; q}\frac{\Theta(\xi_q)}{\xi_q}\frac{1}{\epsilon_0+U_1-U_0-\xi_q}\frac{1}{\epsilon_m+U_1-U_0-\xi_q}$ \\[1ex]
\hline
32 & $c^\dagger_{k'\downarrow} c^\dagger_{q\downarrow}|\o\rangle d^\dagger_{0\uparrow} d_{m\downarrow}|2 M\rangle$ & $c^\dagger_{q\downarrow}|\o\rangle d^\dagger_{0\uparrow}|2 M\rangle$ & $c^\dagger_{q\downarrow} c^\dagger_{k\uparrow}|\o\rangle|2M\rangle$ & same as previous  \\[1ex]
\hline
33 & $c^\dagger_{k'\downarrow} c^\dagger_{k\uparrow}|\o\rangle d^\dagger_{0\uparrow} d_{m\uparrow}|2M\rangle$ & $c^\dagger_{k'\downarrow} c^\dagger_{k\uparrow} c^\dagger_{q\uparrow}|\o\rangle d_{m\uparrow}|2M\rangle$ & $c^\dagger_{k'\downarrow} c^\dagger_{k\uparrow}|\o\rangle|2M\rangle$ & $\frac{1}{\epsilon_0+U_1-U_0}\sum\limits_{m<0 ; q}\frac{\Theta(\xi_q)}{\epsilon_m+\epsilon_0+U_1-U_{-1}-\xi_q}\frac{1}{\epsilon_m+U_1-U_0}$  \\[1ex]
\hline
34 & $c^\dagger_{k'\downarrow} c^\dagger_{k\uparrow}|\o\rangle|2M\rangle$ & $c^\dagger_{k'\downarrow} c^\dagger_{k\uparrow} c^\dagger_{q\downarrow}|\o\rangle d_{m\downarrow}|2M\rangle$ & $c^\dagger_{k'\downarrow} c^\dagger_{k\uparrow}|\o\rangle d^\dagger_{0\downarrow} d_{m\downarrow}|2M\rangle$ & same as previous \\[1ex]
\hline
35 & $c^\dagger_{k'\downarrow} c_{q\downarrow}|\o\rangle d^\dagger_{0\uparrow} d^\dagger_{m\downarrow}|2M\rangle$ & $c_{q\downarrow}|\o\rangle d^\dagger_{0\uparrow} d^\dagger_{0\downarrow} d^\dagger_{m\downarrow}|2M\rangle$ & $|\o\rangle d^\dagger_{0\uparrow} d^\dagger_{0\downarrow}|2M\rangle$ & $\frac{1}{\epsilon_0+U_{2}-U_{1}}\sum\limits_{m>0; q}\frac{\Theta(\xi_q)}{\xi_q+\epsilon_m+U_{2}-U_{1}}\frac{1}{\xi_q+\epsilon_0+\epsilon_m+U_{3}-U_{1}}$   \\[1ex]
\hline
36& $|\o\rangle d^\dagger_{0\uparrow} d^\dagger_{0\downarrow}|2M\rangle$ & $c_{q\uparrow}|\o\rangle d^\dagger_{0\uparrow} d^\dagger_{0\downarrow} d^\dagger_{m\uparrow}|2M\rangle$ & $c_{q\uparrow} c^\dagger_{k\uparrow}|\o\rangle d^\dagger_{0\downarrow} d^\dagger_{m\uparrow}|2M\rangle$ &  same as previous \\[1ex]
\hline\hline
\end{tabular}
\end{table*}

\begin{table*}[t]
\caption{Continuation of Table II}
\vspace{0.1 in}
\begin{tabular}{ c c c c c }
\hline\hline
Label & $|n_1\rangle$ & $|n_2\rangle$ & $|n_3\rangle$ & Contribution to $A^{(4)}_{i\to f}$ (in units of $t^4$)\\
\hline\hline
37 & $c^\dagger_{k'\downarrow} c^\dagger_{q\uparrow}|\o\rangle|2M\rangle$ & $c^\dagger_{k'\downarrow}|\o\rangle d^\dagger_{m\uparrow}|2M\rangle$ & $c^\dagger_{k'\downarrow} c^\dagger_{k\uparrow}|\o\rangle|2M\rangle$ & $-\frac{1}{\epsilon_0+U_1-U_0}\sum\limits_{m>0 ; q}\frac{\Theta(\xi_q)}{\epsilon_0+U_1-U_0-\xi_q}\frac{1}{\epsilon_0-\epsilon_m}$\\[1ex]
\hline
38 & $c^\dagger_{k\uparrow} c^\dagger_{k'\downarrow} |\o\rangle|2M\rangle$ & $c^\dagger_{k\uparrow}|\o\rangle d^\dagger_{m\downarrow}|2M\rangle$ & $c^\dagger_{k\uparrow}c^\dagger_{q\downarrow}|\o\rangle|2M\rangle$ & same as previous \\[1ex]
\hline
39 & $c^\dagger_{k'\downarrow} c_{q\uparrow}|\o\rangle d^\dagger_{0\uparrow} d^\dagger_{m\uparrow}|2M\rangle$ & $c^\dagger_{k'\downarrow}|\o\rangle d^\dagger_{m\uparrow}|2M\rangle$ & $c^\dagger_{k'\downarrow} c^\dagger_{k\uparrow}|\o\rangle|2M\rangle$ & $\frac{1}{\epsilon_0+U_1-U_0}\sum\limits_{m>0; q}\frac{\Theta(\xi_q)}{-\epsilon_m+U_1-U_2-\xi_q}\frac{1}{\epsilon_0-\epsilon_m}$\\[1ex]
\hline
40 & $c^\dagger_{k'\downarrow} c^\dagger_{k\uparrow}|\o\rangle|2M\rangle$ & $c^\dagger_{k\uparrow}|\o\rangle d^\dagger_{m\downarrow}|2M\rangle$ & $c^\dagger_{k\uparrow} c_{q\downarrow}|\o\rangle d^\dagger_{0\downarrow} d^\dagger_{m\downarrow}|2M\rangle$ & same as previous \\[1ex]
\hline
41 & $c^\dagger_{k'\downarrow} c^\dagger_{q\downarrow}|\o\rangle d^\dagger_{0\uparrow} d_{m\downarrow}|2M\rangle$ & $c^\dagger_{k'\downarrow}|\o\rangle d^\dagger_{0\uparrow} d^\dagger_{0\downarrow} d_{m\downarrow}|2M\rangle$ & $c^\dagger_{k'\downarrow} c^\dagger_{k\uparrow}|\o\rangle d^\dagger_{0\downarrow} d_{m\downarrow}|2M\rangle$ & $\sum\limits_{m<0;q}\frac{1}{\epsilon_m-\epsilon_0}\frac{1}{\epsilon_m+U_1-U_0}\frac{\Theta(\xi_q)}{\epsilon_m+U_1-U_0-\xi_q}$\\[1ex]
\hline
42 & $c^\dagger_{k\uparrow} c^\dagger_{k'\downarrow}|\o\rangle d^\dagger_{0\uparrow} d_{m\uparrow}|2M\rangle$ & $c^\dagger_{k\uparrow}|\o\rangle d^\dagger_{0\uparrow} d^\dagger_{0\downarrow} d_{m\uparrow}|2M\rangle$ & $c^\dagger_{k\uparrow} c^\dagger_{q\uparrow}|\o\rangle d^\dagger_{0\downarrow} d_{m\uparrow}|2M\rangle$ & same as previous \\[1ex]
\hline
43 & $c^\dagger_{q\uparrow} c^\dagger_{k'\downarrow}|\o\rangle|2 M\rangle$ & $c^\dagger_{k'\downarrow}|\o\rangle d^\dagger_{m\uparrow}|2M\rangle$ & $|\o\rangle d^\dagger_{0\downarrow} d^\dagger_{m\uparrow}|2M\rangle$ & $-\sum\limits_{m>0 ; q}\frac{1}{-\epsilon_m+U_1-U_2}\frac{1}{\epsilon_0-\epsilon_m}\frac{\Theta(\xi_q)}{\epsilon_0+U_1-U_0-\xi_q}$\\[1ex]
\hline
44 & $|\o\rangle d^\dagger_{0\uparrow} d^\dagger_{m\downarrow}|2M\rangle$ & $c^\dagger_{k\uparrow}|\o\rangle d^\dagger_{m\downarrow}|2M\rangle$ & $c^\dagger_{k\uparrow} c^\dagger_{q\downarrow}|\o\rangle|2M\rangle$ & same as previous \\[1ex]
\hline
45 & $c^\dagger_{k'\downarrow} c_{q\downarrow}|\o\rangle d^\dagger_{0\uparrow} d^\dagger_{0\downarrow}|2M\rangle$ & $c^\dagger_{k'\downarrow}|\o\rangle d^\dagger_{0\uparrow} d^\dagger_{0\downarrow} d_{m\downarrow}|2M\rangle$ & $c^\dagger_{k'\downarrow} c^\dagger_{k\uparrow}|\o\rangle d^\dagger_{0\downarrow} d_{m\downarrow}|2M\rangle$ & $-\sum\limits_{m<0 ; q}\frac{1}{\epsilon_m-\epsilon_0}\frac{1}{\epsilon_m+U_1-U_0}\frac{\Theta(\xi_q)}{-\epsilon_0+U_1-U_2-\xi_q}$\\[1ex]
\hline
46 & $c^\dagger_{k\uparrow} c^\dagger_{k'\downarrow}|\o\rangle d^\dagger_{0\uparrow} d_{m\uparrow}|2M\rangle$ & $c^\dagger_{k\uparrow}|\o\rangle d^\dagger_{0\uparrow} d^\dagger_{0\downarrow} d_{m\uparrow}|2M\rangle$& $c^\dagger_{k\uparrow} c_{q\uparrow}|\o\rangle d^\dagger_{0\uparrow} d^\dagger_{0\downarrow}|2M\rangle$ & same as previous \\[1ex]
\hline
47 & $c^\dagger_{k'\downarrow} c_{q\uparrow}|\o\rangle d^\dagger_{0\uparrow} d^\dagger_{m\uparrow}|2M\rangle$ & $c^\dagger_{k'\downarrow}|\o\rangle d^\dagger_{m\uparrow}|2M\rangle$ & $|\o\rangle d^\dagger_{m\uparrow} d^\dagger_{0\downarrow}|2M\rangle$ & $\sum\limits_{m>0; q}\frac{1}{-\epsilon_m+U_1-U_2}\frac{1}{\epsilon_0-\epsilon_m}\frac{\Theta(\xi_q)}{-\epsilon_m+U_1-U_2-\xi_q}$\\[1ex]
\hline
48 & $|\o\rangle d^\dagger_{m\downarrow} d^\dagger_{0\uparrow}|2M\rangle$ & $c^\dagger_{k\uparrow}|\o\rangle d^\dagger_{m\downarrow}|2M\rangle$ & $c^\dagger_{k\uparrow} c_{q\downarrow}|\o\rangle d^\dagger_{0\downarrow} d^\dagger_{m\downarrow}|2M\rangle$ & same as previous \\[1ex]
\hline
49 & $c^\dagger_{k'\downarrow} c_{q\downarrow}|\o\rangle d^\dagger_{0\uparrow} d^\dagger_{0\downarrow}|2M\rangle$ & $c^\dagger_{k'\downarrow}|\o\rangle d^\dagger_{0\uparrow} d^\dagger_{0\downarrow} d_{m\downarrow}|2M\rangle$ & $|\o\rangle d^\dagger_{0\uparrow} d^\dagger_{0\downarrow}|2M\rangle$ & $-\frac{1}{-\epsilon_0+U_1-U_2}\sum\limits_{m<0; q}\frac{1}{\epsilon_m-\epsilon_0}\sum_q\frac{\Theta(\xi_q)}{-\epsilon_0+U_1-U_2-\xi_q}$ \\[1ex]
\hline
50 & $|\o\rangle d^\dagger_{0\uparrow} d^\dagger_{0\downarrow}|2M\rangle$ & $c^\dagger_{k\uparrow}|\o\rangle d^\dagger_{0\uparrow} d^\dagger_{0\downarrow} d_{m\uparrow} |2M\rangle$ & $c^\dagger_{k\uparrow} c_{q\uparrow}|\o\rangle d^\dagger_{0\uparrow} d^\dagger_{0\downarrow}|2M\rangle$ & same as previous \\[1ex]
\hline
51 & $c^\dagger_{k'\downarrow} c^\dagger_{q\downarrow}|\o\rangle d^\dagger_{0\uparrow} d_{m\downarrow} |2M\rangle$ & $c^\dagger_{k'\downarrow}|\o\rangle d^\dagger_{0\uparrow} d^\dagger_{0\downarrow} d_{m\downarrow}|2M\rangle$ & $|\o\rangle d^\dagger_{0\uparrow} d^\dagger_{0\downarrow} |2M\rangle$ & $\frac{1}{-\epsilon_0+U_1-U_2}\sum\limits_{m<0 ; q}\frac{1}{\epsilon_m-\epsilon_0}\frac{\Theta(\xi_q)}{\epsilon_m+U_1-U_0-\xi_q}$\\[1ex]
\hline
52 & $|\o\rangle d^\dagger_{0\uparrow} d^\dagger_{0\downarrow} |2M\rangle$ & $c^\dagger_{k\uparrow}|\o\rangle d^\dagger_{0\uparrow} d^\dagger_{0\downarrow} d_{m\uparrow}|2M\rangle$ & $c^\dagger_{k\uparrow} c^\dagger_{q\uparrow}|\o\rangle d^\dagger_{0\downarrow} d_{m\uparrow}|2M\rangle$ & same as previous \\[1ex]
\hline
53 & $c^\dagger_{k'\downarrow} c^\dagger_{q\uparrow}|\o\rangle d^\dagger_{0\uparrow} d_{m\uparrow}|2M\rangle$ & $c^\dagger_{k'\downarrow} c^\dagger_{k\uparrow} c^\dagger_{q\uparrow} |\o\rangle d_{m\uparrow}|2 M\rangle$ & $c^\dagger_{k'\downarrow} c^\dagger_{k\uparrow} |\o\rangle |2M\rangle$ & $-\frac{1}{\epsilon_0+U_1-U_0}\sum\limits_{m<0; q}\frac{\Theta(\xi_q)}{\epsilon_m+U_1-U_0-\xi_q}\frac{1}{\epsilon_m+\epsilon_0+U_1-U_{-1}-\xi_q}$\\[1ex]
\hline
54 & $c^\dagger_{k'\downarrow} c^\dagger_{k\uparrow}|\o\rangle|2M\rangle$ & $c^\dagger_{k'\downarrow} c^\dagger_{k\uparrow} c^\dagger_{q\downarrow}|\o\rangle d_{m\downarrow}|2M\rangle$ & $c^\dagger_{k\uparrow} c^\dagger_{q\downarrow}|\o\rangle d^\dagger_{0\downarrow} d_{m\downarrow}|2M\rangle$ & same as previous\\[1ex]
\hline
55 & $c^\dagger_{k'\downarrow} c_{q\uparrow}|\o\rangle d^\dagger_{0\uparrow} d^\dagger_{m\uparrow}|2M\rangle$ & $c^\dagger_{k'\downarrow} c^\dagger_{k\uparrow} c_{q\uparrow}|\o\rangle d^\dagger_{m\uparrow}|2M\rangle$ & $c^\dagger_{k'\downarrow} c^\dagger_{k\uparrow} |\o\rangle|2M\rangle$ & same as No. 7\\[1ex]
\hline
56 & $c^\dagger_{k'\downarrow} c^\dagger_{k\uparrow}|\o\rangle|2M\rangle$ & $c^\dagger_{k'\downarrow} c^\dagger_{k\uparrow} c_{q\downarrow}|\o\rangle d^\dagger_{m\downarrow}|2M\rangle$ & $c^\dagger_{k\uparrow} c_{q\downarrow}|\o\rangle d^\dagger_{0\downarrow} d^\dagger_{m\downarrow}|2M\rangle$ & same as previous\\[1ex]
\hline
57 & $|\o\rangle d^\dagger_{0\uparrow} d^\dagger_{0\downarrow}|2M\rangle$ & $c^\dagger_{q\uparrow}|\o\rangle d^\dagger_{0\uparrow} d^\dagger_{0\downarrow} d_{m\uparrow} |2M\rangle$ & $c^\dagger_{q\uparrow} c^\dagger_{k\uparrow}|\o\rangle d^\dagger_{0\downarrow} d_{m\uparrow}|2M\rangle$ & $-\frac{1}{-\epsilon_0+U_1-U_2}\sum\limits_{m<0; q}\frac{\Theta(\xi_q)}{\epsilon_m-\epsilon_0-\xi_q}\frac{1}{\epsilon_m+U_1-U_0-\xi_q}$\\[1ex]
\hline
58 & $c^\dagger_{q\downarrow} c^\dagger_{k'\downarrow}|\o\rangle d^\dagger_{0\uparrow} d_{m\downarrow}|2M\rangle$ & $c^\dagger_{q\downarrow}|\o\rangle d^\dagger_{0\uparrow} d^\dagger_{0\downarrow}d_{m\downarrow}|2M\rangle$ & $|\o\rangle d^\dagger_{0\uparrow} d^\dagger_{0\downarrow}|2M\rangle$ & same as previous\\[1ex]
\hline
59 & $c^\dagger_{k'\downarrow} c_{q\uparrow}|\o\rangle d^\dagger_{0\uparrow} d^\dagger_{m\uparrow}|2M\rangle$ & $c_{q\uparrow}|\o\rangle d^\dagger_{0\uparrow} d^\dagger_{0\downarrow} d^\dagger_{m\uparrow}|2M\rangle$ & $|\o\rangle d^\dagger_{0\uparrow} d^\dagger_{0\downarrow}|2M\rangle$ & same as No. 35\\[1ex] 
\hline
60 & $|\o\rangle d^\dagger_{0\uparrow} d^\dagger_{0\downarrow}|2M\rangle$ & $c_{q\downarrow}|\o\rangle d^\dagger_{0\uparrow} d^\dagger_{0\downarrow} d^\dagger_{m\downarrow}|2M\rangle$ & $c^\dagger_{k\uparrow} c_{q\downarrow}|\o\rangle d^\dagger_{0\downarrow} d^\dagger_{m\downarrow}|2M\rangle$ & same as previous\\[1ex]
\hline
61 & $c^\dagger_{q\downarrow} c^\dagger_{k'\downarrow}|\o\rangle d_{m\downarrow} d^\dagger_{0\uparrow}|2M\rangle$ & $c^\dagger_{k\uparrow} c^\dagger_{k'\downarrow} c^\dagger_{q\downarrow}|\o\rangle d_{m\downarrow}|2M\rangle$ & $c^\dagger_{k\uparrow} c^\dagger_{k'\downarrow}|\o\rangle|2M\rangle$ & same as No. 53\\[1ex]
\hline
62 & $c^\dagger_{k'\downarrow} c^\dagger_{k\uparrow}|\o\rangle|2M\rangle$ & $c^\dagger_{k\uparrow} c^\dagger_{k'\downarrow} c^\dagger_{q\uparrow}|\o\rangle d_{m\uparrow}|2M\rangle$ & $c^\dagger_{k\uparrow} c^\dagger_{q\uparrow}|\o\rangle d_{m\uparrow} d^\dagger_{0\downarrow}|2M\rangle$ &  same as previous\\[1ex]
\hline
63 & $|\o\rangle d^\dagger_{0\uparrow} d^\dagger_{0\downarrow}|2M\rangle$ & $c^\dagger_{q\downarrow}|\o\rangle d^\dagger_{0\uparrow} d^\dagger_{0\downarrow} d_{m\downarrow}|2M\rangle$ & $c^\dagger_{q\downarrow} c^\dagger_{k\uparrow}|\o\rangle d^\dagger_{0\downarrow} d_{m\downarrow}|2M\rangle$ & same as No. 57\\[1ex]
\hline
64 & $c^\dagger_{q\uparrow} c^\dagger_{k'\downarrow}|\o\rangle d^\dagger_{0\uparrow} d_{m\uparrow} |2M\rangle$ & $c^\dagger_{q\uparrow}|\o\rangle d^\dagger_{0\uparrow} d^\dagger_{0\downarrow} d_{m\uparrow}|2M\rangle$ & $|\o\rangle d^\dagger_{0\uparrow} d^\dagger_{0\downarrow}|2 M\rangle$ & same as previous\\[1ex]
\hline\hline
\end{tabular}
\end{table*}

\begin{table*}[t]
\caption{Continuation of Table III. Unlike in Tables II and III, every intermediate configuration in this Table produces a divergence-free amplitude.}
\vspace{0.1 in}
\begin{tabular}{ c c c c c }
\hline\hline
Label & $|n_1\rangle$ & $|n_2\rangle$ & $|n_3\rangle$ & Contribution to $A^{(4)}_{i\to f}$ (in units of $t^4$)\\
\hline\hline
65 & $|\o\rangle d^\dagger_{0\uparrow} d^\dagger_{m\downarrow} |2M\rangle$ & $c_{q\downarrow}|\o\rangle d^\dagger_{0\uparrow} d^\dagger_{0\downarrow} d^\dagger_{m\downarrow} |2M\rangle$ & $c_{q\downarrow} c^\dagger_{k\uparrow}|\o\rangle d^\dagger_{0\downarrow} d^\dagger_{m\downarrow}|2M\rangle$ & $\sum\limits_{q;m>0}\frac{\Theta(\xi_q)}{-\epsilon_m+U_1-U_2}\frac{1}{-\epsilon_0-\epsilon_m-\xi_q+U_1-U_3}\frac{1}{-\epsilon_m-\xi_q+U_1-U_2}$  \\[1ex]
\hline
66 & $c^\dagger_{q\uparrow} c^\dagger_{k'\downarrow}|\o\rangle|2M\rangle$ & $c^\dagger_{k\uparrow} c^\dagger_{k'\downarrow} c^\dagger_{q\uparrow}|\o\rangle d_{m\uparrow}|2M\rangle$ & $c^\dagger_{k\uparrow} c^\dagger_{q\uparrow}|\o\rangle d^\dagger_{0\downarrow} d_{m\uparrow}|2M\rangle$ & $\sum\limits_{m<0;q}\frac{\Theta(\xi_q)}{\epsilon_0-\epsilon_q+U_1-U_0}\frac{1}{\epsilon_0+\epsilon_m-\xi_q+U_1-U_{-1}}\frac{1}{\epsilon_m-\xi_q+U_1-U_0} $\\[1ex]
\hline
67 & $c^\dagger_{k'\downarrow} c_{q\downarrow}|\o\rangle d^\dagger_{0\uparrow} d^\dagger_{m\downarrow}|2M\rangle$ & $c_{q\downarrow}|\o\rangle d^\dagger_{0\uparrow} d^\dagger_{0\downarrow} d^\dagger_{m\downarrow} |2M\rangle$ & $c_{q\downarrow} c^\dagger_{k\uparrow}|\o\rangle d^\dagger_{0\downarrow} d^\dagger_{m\downarrow}|2M\rangle$ & $-\sum\limits_{q;m>0} \frac{\Theta(\xi_q)}{-\epsilon_m-\xi_q+U_1-U_2}\frac{1}{-\epsilon_0-\epsilon_m-\xi_q+U_1-U_3}\frac{1}{-\epsilon_m-\xi_q+U_1-U_2}$\\[1ex]
\hline
68 & $c^\dagger_{k'\downarrow} c_{q\downarrow}|\o\rangle d^\dagger_{0\uparrow} d^\dagger_{m\downarrow}|2M\rangle$ & $c^\dagger_{k\uparrow}c^\dagger_{k'\downarrow} c_{q\downarrow}|\o\rangle d^\dagger_{m\downarrow}|2M\rangle$ & $c_{q\downarrow} c^\dagger_{k\uparrow}|\o\rangle d^\dagger_{0\downarrow} d^\dagger_{m\downarrow}|2M\rangle$ & $-\sum\limits_{q;m>0}\frac{\Theta(\xi_q)}{-\epsilon_m-\xi_q+U_1-U_2}\frac{1}{\epsilon_0-\epsilon_m-\xi_q}\frac{1}{-\epsilon_m-\xi_q+U_1-U_2}$\\[1ex]
\hline
69 & $c^\dagger_{k'\downarrow} c_{q\downarrow}|\o\rangle d^\dagger_{0\uparrow} d^\dagger_{0\downarrow}|2M\rangle$ & $c_{q\downarrow}|\o\rangle d^\dagger_{0\uparrow} d^\dagger_{0\downarrow} d^\dagger_{m\downarrow}|2M\rangle$ & $c_{q\downarrow} c^\dagger_{k\uparrow}|\o\rangle d^\dagger_{0\downarrow} d^\dagger_{m\downarrow}|2M\rangle$ & $\frac{\Theta(\xi_q)}{-\epsilon_0-\epsilon_q+U_1-U_2}\frac{1}{-\epsilon_0-\epsilon_m-\xi_q+U_1-U_3}\frac{1}{-\epsilon_m-\epsilon_q+U_1-U_2}$\\[1ex]
\hline
70 & $c^\dagger_{k\uparrow} c^\dagger_{k'\downarrow}|\o\rangle d^\dagger_{0\uparrow} d_{m\uparrow}|2M\rangle$ & $c^\dagger_{k'\downarrow} c^\dagger_{k\uparrow} c^\dagger_{q\uparrow}|\o\rangle d_{m\uparrow}|2 M\rangle$ & $c^\dagger_{k\uparrow} c^\dagger_{q\uparrow}|\o\rangle d^\dagger_{0\downarrow} d_{m\uparrow}|2M\rangle$ & $\sum\limits_{m<0;q} \frac{\Theta(\xi_q)}{\epsilon_m+U_1-U_0}\frac{1}{\epsilon_0+\epsilon_m-\xi_q+U_1-U_{-1}}\frac{1}{\epsilon_m-\xi_q+U_1-U_0}$   \\[1ex]
\hline
71 & $c^\dagger_{k'\downarrow} c^\dagger_{q\downarrow}|\o\rangle d^\dagger_{0\uparrow} d_{m\downarrow}|2M\rangle$ & $c^\dagger_{k\uparrow} c^\dagger_{k'\downarrow} c^\dagger_{q\downarrow}|\o\rangle d_{m\downarrow}|2M\rangle$ & $c^\dagger_{k\uparrow} c^\dagger_{k'\downarrow}|\o\rangle d^\dagger_{0\downarrow} d_{m\downarrow}|2M\rangle$ & Same as No. 70\\[1ex]
\hline
72 & $c^\dagger_{k'\downarrow} c^\dagger_{q\downarrow}|\o\rangle d^\dagger_{0\uparrow} d_{m\downarrow}|2M\rangle$ & $c^\dagger_{k\uparrow} c^\dagger_{k'\downarrow} c^\dagger_{q\downarrow}|\o\rangle d_{m\downarrow}|2M\rangle$ & $c^\dagger_{k\uparrow} c^\dagger_{q\downarrow}|\o\rangle|2M\rangle$ & Same as No. 66\\[1ex]
\hline
73 & $c^\dagger_{k'\downarrow} c^\dagger_{q\downarrow}|\o\rangle d^\dagger_{0\uparrow} d_{m\downarrow}|2M\rangle$ & $c^\dagger_{k\uparrow} c^\dagger_{k'\downarrow} c^\dagger_{q\downarrow}|\o\rangle d_{m\downarrow}|2M\rangle$ & $c^\dagger_{k\uparrow} c^\dagger_{q\downarrow}|\o\rangle d^\dagger_{0\downarrow} d_{m\downarrow}|2M\rangle$ & $-\sum\limits_{m<0;q} \frac{\Theta(\xi_q)}{\epsilon_m-\xi_q+U_1-U_0}\frac{1}{\epsilon_0+\epsilon_m-\xi_q+U_1-U_{-1}}\frac{1}{\epsilon_m-\xi_q+U_1-U_0}$    \\[1ex]
\hline
74 & $c^\dagger_{k'\downarrow} c^\dagger_{q\downarrow}|\o\rangle d^\dagger_{0\uparrow} d_{m\downarrow}|2M\rangle$ & $c^\dagger_{q\downarrow}|\o\rangle d^\dagger_{0\uparrow} d^\dagger_{0\downarrow} d_{m\downarrow}|2M\rangle$ & $c^\dagger_{q\downarrow} c^\dagger_{k\uparrow}|\o\rangle d^\dagger_{0\downarrow} d_{m\downarrow}|2M\rangle$ & $-\sum\limits_{m<0;q}\frac{\Theta(\xi_q)}{\epsilon_m-\xi_q+U_1-U_0}\frac{1}{-\epsilon_0+\epsilon_m-\xi_q}\frac{1}{\epsilon_m-\xi_q+U_1-U_0}$    \\[1ex]
\hline
75 & $c^\dagger_{k'\downarrow} c_{q\uparrow}|\o\rangle d^\dagger_{0\uparrow} d^\dagger_{m\uparrow}|2M\rangle$ & $c_{q\uparrow}|\o\rangle d^\dagger_{0\uparrow} d^\dagger_{0\downarrow} d^\dagger_{m\uparrow}|2M\rangle$ & $c_{q\uparrow} c^\dagger_{k\uparrow}|\o\rangle d^\dagger_{0\uparrow} d^\dagger_{0\downarrow}|2M\rangle$ & Same as No. 69\\[1ex]
\hline
76 & $c^\dagger_{k'\downarrow} c_{q\uparrow}|\o\rangle d^\dagger_{0\uparrow} d^\dagger_{m\uparrow}|2M\rangle$ & $c_{q\uparrow}|\o\rangle d^\dagger_{0\uparrow} d^\dagger_{0\downarrow} d^\dagger_{m\uparrow}|2M\rangle$ & $c_{q\uparrow} c^\dagger_{k\uparrow}|\o\rangle d^\dagger_{m\uparrow} d^\dagger_{0\downarrow}|2M\rangle$ & Same as No. 67\\[1ex]
\hline
77 & $c^\dagger_{k'\downarrow} c_{q\uparrow}|\o\rangle d^\dagger_{0\uparrow} d^\dagger_{m\uparrow}|2M\rangle$ & $c_{q\uparrow}|\o\rangle d^\dagger_{0\uparrow} d^\dagger_{0\downarrow} d^\dagger_{m\uparrow}|2M\rangle$ & $|\o\rangle d^\dagger_{0\downarrow} d^\dagger_{m\uparrow}|2M\rangle$ & Same as No. 65\\[1ex]
\hline
78 & $c^\dagger_{k'\downarrow} c_{q\uparrow}|\o\rangle d^\dagger_{0\uparrow} d^\dagger_{m\uparrow}|2M\rangle$ & $c^\dagger_{k'\downarrow} c^\dagger_{k\uparrow} c_{q\uparrow}|\o\rangle d^\dagger_{m\uparrow}|2M\rangle$ & $c^\dagger_{k\uparrow} c_{q\uparrow}|\o\rangle d^\dagger_{0\downarrow} d^\dagger_{m\uparrow}|2M\rangle$ & Same as No. 68\\[1ex]
\hline
79 & $c^\dagger_{k'\downarrow} c^\dagger_{q\uparrow}|\o\rangle d^\dagger_{0\uparrow} d_{m\uparrow}|2M\rangle$ & $c^\dagger_{k\uparrow} c^\dagger_{k'\downarrow} c^\dagger_{q\uparrow}|\o\rangle d_{m\uparrow}|2M\rangle$ & $c^\dagger_{k\uparrow} c^\dagger_{q\uparrow}|\o\rangle d^\dagger_{0\downarrow} d_{m\uparrow}|2M\rangle$ & Same as No. 73\\[1ex]
\hline
80 & $c^\dagger_{k'\downarrow} c^\dagger_{q\uparrow}|\o\rangle d^\dagger_{0\uparrow} d_{m\uparrow}|2M\rangle$ & $c^\dagger_{q\uparrow}|\o\rangle d^\dagger_{0\uparrow} d^\dagger_{0\downarrow} d_{m\uparrow}|2M\rangle$ & $c^\dagger_{q\uparrow} c^\dagger_{k\uparrow}|\o\rangle d^\dagger_{0\downarrow} d_{m\uparrow}|2M\rangle$ & Same as No. 74\\[1ex]
\hline\hline
\end{tabular}
\end{table*}

\end{widetext}

\begin{thebibliography}{50}
\bibitem{coleman} For a review see e.g. P. Coleman, {\em Many-Body Physics}, http://www.physics.rutgers.edu/$\sim$coleman/mbody.html.
\bibitem{haldane1978} F.D.M. Haldane, J. Phys. C {\bf 11}, 5015 (1978).
\bibitem{kondo exp} D. Goldhaber-Gordon {\em et al.}, Nature {\bf 391}, 156 (1998); S.M. Cronenwett {\em et al.}, Science {\bf 281}, 540 (1998).
\bibitem{inoshita1993} T. Inoshita {\em et al.}, Phys. Rev. B {\bf 48}, 14725 (1993).
\bibitem{kondo exp 2} D. Goldhaber-Gordon {\em et al.}, Phys. Rev. Lett. {\bf 81}, 5225 (1998); W.G. van der Wiel {\em et al.}, Science {\bf 289}, 2105 (2000).
\bibitem{aleiner2002} I.L. Aleiner {\em et al.}, Phys. Rep. {\bf 358}, 309 (2002).
\bibitem{suhl1965} H. Suhl, Phys. Rev. {\bf 138}, A515 (1965); H. Suhl in {\em Theory of Magnetism in Transition Metals}, ed. W. Marshall (Academic Press, New York, 1967).
\bibitem{messiah} See e.g. A. Messiah, {\em Quantum Mechanics (vol. II)} (North-Holland Publishing Co., Amsterdam, 1962). 
\bibitem{langreth1966} D.C. Langreth, Phys. Rev. {\bf 150}, 516 (1966).
\bibitem{sorensen1996} E.S. Sorensen and I. Affleck, Phys. Rev. B {\bf 53}, 9153 (1996).
\bibitem{sakurai} J.J. Sakurai, {\em Modern Quantum Mechanics} (Addison-Wesley, Reading, MA, 1994).
\end{thebibliography}
\end{document}